\documentstyle[12pt,epsfig,pennames]{article}
\def\D{\Delta}

\def\f{\phi}

\def\c{\xi}

\def\p{\pi}

\def\r{\rho}

\def\DpT{\Delta p_\rT}

\def\Ncc{N_{\rc\rc}}

\def\vs{\vskip}

\def\ifmath#1{\relax\ifmmode #1\else $#1$\fi}%
\def\rd{\ifmath{{\mathrm{d}}}}
\def\rD{\ifmath{{\mathrm{D}}}}

\def\rM{\ifmath{{\mathrm{M}}}}
\def\rT{\ifmath{{\mathrm{T}}}}
\def\rc{\ifmath{{\mathrm{c}}}}

\def\norm{\ifmath{{\mathrm{norm}}}}

\def\vec#1{{\mbox{\bf #1}}}
 
\newcommand{\beqa}{\begin{eqnarray}}
\newcommand{\eeqa}{\end{eqnarray}  }
\newcommand{\beqan}{\begin{eqnarray*}}
\newcommand{\eeqan}{\end{eqnarray*}}
\newcommand{\beq}{\begin{equation}}
\newcommand{\eeq}{\end{equation}  }
 
\fussy

\begin{document}
\begin{titlepage}
\pagestyle{empty}
\vspace*{4cm}
\begin{center}
{\large\bf Correlations in transverse momentum \\
in $\p^+\Pp$ and $\PK^+\Pp$ collisions at 250 GeV/$c$\footnote{Talk
presented at the 7th Workshop on Multiparticle Production ``Correlations
and Fluctuations'', Nijmegen, The Netherlands, 30th June--6th July, 1996}}
\vspace{1.1cm}\\
{\sc J.~Czy\.zewski}\footnote{Fellow of the Polish 
             Science Foundation (FNP) scholarship for the year 1996}
\\
{\it Institute of Physics, Jagellonian University,\\ 
ul.\ Reymonta 4, PL-30-059 Krak\'ow, Poland}
\vspace{0.4cm}\\
for EHS/NA22 Collaboration
\end{center}
\vspace{0.0cm}
\begin{abstract}  
We have measured the second-order normalized differential factorial moments 
as a function of the difference of 
transverse momentum ($\Delta p_\rT$) in $\p^+\Pp$ and $\PK^+\Pp$ collisions
at 250 GeV/$c$. The second-order differential factorial moments 
for like-charged pairs reveal a strong increase with decreasing $\Delta p_\rT$. 
In a small central rapidity window this increase is described 
by a simple power law. Such a behavior, if interpreted as 
originating from Bose-Einstein correlations, may indicate a structure of 
the transverse spatial distribution of the source similar to that recently 
predicted by Bia\l{}as and Peschanski for color-dipole emission in 
onium-onium scattering. The power of the rise obtained in the fit agrees
with the predicted value. 
\end{abstract} 
\vspace{0.5cm}

\noindent
{\sf TPJU 24/96}\\
{\sf December 1996}
\end{titlepage}
%
%
 
\section{Introduction}
It has recently been shown by Bia\l{}as and Peschanski$^1$ that
the cross-section for the emission of color dipoles in high-energy 
onium-onium scattering reveals a power law in transverse position of the 
emitted dipole. The authors of Ref.~1 use the formalism recently 
developed by Mueller$^2$ in which an onium state resembles a 
collection of color dipoles of various sizes. The authors calculated the 
contribution to the cross-section from the emission of dipoles present in the 
initial state and released during the collision. This cross-section depends on
the transverse position $r_\rT$ of the dipole as
\beq
{\rd \sigma \over \rd x \rd^2 r_\rT} \sim r_\rT^{-2 + \gamma_\rM},
\label{power}
\eeq
with $x$ being the transverse spatial size of the dipole and
$\gamma_\rM \approx 0.37$. The quoted value of $\gamma_\rM$ does not 
depend on any physical parameters of the model, such as the strong coupling
constant, the masses involved, etc. It is governed only by some very 
general properties of the underlying cascade.

The power law (\ref{power}) follows from the self-similar
character of gluon emission in the onium state and leads to a rise of the 
factorial moments$^3$ with decreasing distance in transverse 
momentum between the emitted particles. 

The exact shape of the 
transverse-position cross-section is also related to the Bose-Einstein 
correlation between identical particles$^4$ as measured in the 
transverse momentum difference. 
Since the latter is the square of the Fourier transform of the 
source distribution, a transverse-position cross-section of the form
(\ref{power}) implies that the correlation depends on the absolute
value of the transverse-momentum difference $\DpT$ like 
$\Delta p_\rT^{-2\gamma_\rM}$. Due to cut-offs in the cross-section 
(\ref{power}) necessary for its normalization, this power-law behavior of 
the correlation is restricted to a certain region in 
$\Delta p_\rT$$^{1,5}$.

The result (\ref{power}) of Ref.~1 was derived
in perturbative QCD and one could argue that it holds only for virtually
infinite energies. Nevertheless, in the derivation of Eq.~\ref{power}
one operates only on color dipoles, which are colorless objects and interact
only by colorless two-gluon exchange. One can expect thus that such 
an approach to a very-high-energy collision can describe to a good degree
many phenomena occurring with hadrons (which are also colorless objects)
at finite energies. This is not the case for the usual perturbative
QCD calculations,
which describe the dynamics of colored objects (quarks and gluons).

Thus, in spite of the fact that the transverse-position cross-section
was derived for onium-onium scattering at an asymptotically high energy,
it is tempting to verify whether there is any indication of such a behavior
in normal hadron-hadron scattering. This verification will be described
in the present paper for $\pi^+\Pp$ and $\PK^+\Pp$ collisions at 250GeV/$c$.

\section{Measured quantities}
For measuring the two-particle correlations, we used the differential
normalized factorial moments evaluated using the density integrals$^6$
\beq
\rD F_2(\DpT) = {1\over \Ncc}{\rD f_2(\DpT) \over \rD\c^{\norm}_2(\DpT)}
\label{DF2}
\eeq
with
\beq
\rD f_2(\DpT)=\int \rho_2(\vec{p}_1, \vec{p}_2)\delta_{12} \rd^3\vec{p}_1 
\rd^3\vec{p}_2
\eeq
and $\delta_{12}$ defined to be 1 when $|\vec{p}_{\rT1}-\vec{p}_{\rT2}|$ lies
within a certain bin around $\DpT$ and 0 otherwise.
$\rD\c^{\norm}_2$ is defined by the integral
\beq
\rD\c^{\norm}_2(\DpT) = \int \r_1(\vec{p}_1) \r_1(\vec{p}_2) \delta_{12} 
\rd^3\vec{p}_1 \rd^3\vec{p}_2
\eeq
evaluated with particles 1 and 2 taken randomly from different events 
(``event mixing"). The property of unbiased estimators for the moments and 
their normalization is demonstrated in Ref.~7. The non-trivial 
modifications needed for events with non-uniform weights are derived 
in Ref.~8.

The normalization factor $\Ncc$ depends on the charge combination of the
measured pairs: $\Ncc = \langle n(n-1)\rangle/\langle n\rangle^2$,
$N_{+-} = \langle n_+n_-\rangle/\langle n_+\rangle\langle n_-\rangle$,
$N_{--} = \langle n_-(n_--1)\rangle/\langle n_-\rangle^2$,
$N_{++} = \langle n_+(n_+-1)\rangle/\langle n_+\rangle^2$
for all pairs, unlike-charge pairs, negative-negative, and 
positive-positive pairs, respectively. The symbols 
$n$, $n_-$, and $n_+$ stand for the numbers of total, negative,
and positive particles, respectively, and $\langle\rangle$ stands for
averaging over all events in the data sample used.
Differential factorial moments normalized by the factor $\Ncc$ are
equal to 1 when there is no correlation between the particles.

On the other hand, the normalized differential factorial moments are equivalent
to the disconnected correlation function measured at a given value
of $\DpT$, but integrated over the difference in the longitudinal
momentum.

\section{Results}
For this CERN experiment, the European Hybrid Spectrometer (EHS) was
equipped with the Rapid Cycling Bubble Chamber (RCBC) as an active  vertex
detector and exposed to a 250 GeV/$c$ tagged positive, meson-enriched beam.
In data taking, a minimum-bias interaction trigger was used. The
details of the spectrometer and the trigger
can be found in previous publications$^{9,10}$.

For the reported analysis the inelastic non-single-diffractive
sample consisting of 59~200 $\p^+\Pp$ and $\PK^+\Pp$ events was used.

\begin{figure}
\begin{center}
\mbox{\epsfig{file=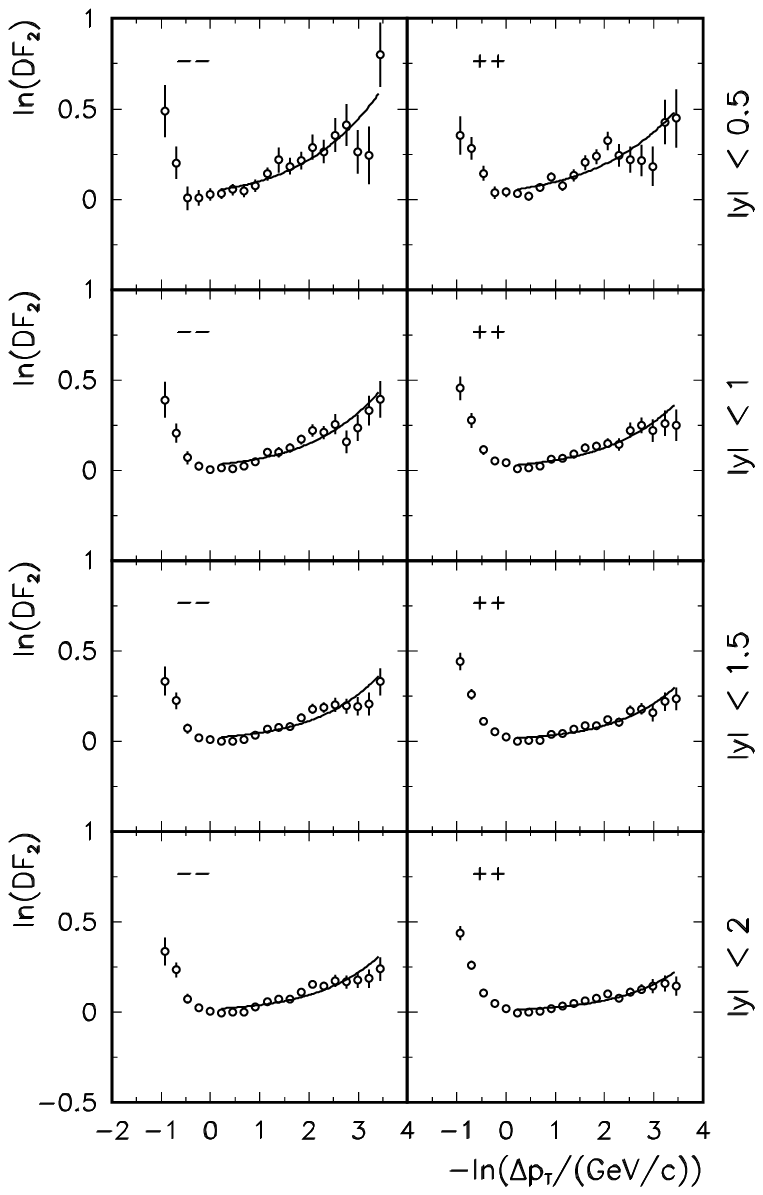,width=12.5cm}}
\parbox{4.8in}{
\small Figure 1: Logarithm of the second-order differential 
factorial moment $\rD F_2$ as a function of $-\ln(\DpT/($GeV$/c))$ 
for four different rapidity cuts. The bottom plots correspond to the 
full data sample. The solid lines represent the fit by Eq.~\ref{powerfit}.
}
\end{center}
\end{figure}

In Fig.~1 we show plots of the second-order differential factorial moments $\rD
F_2$ obtained according to Eq.~\ref{DF2} for the negative-negative ($--$) 
pairs and for the positive-positive (++) ones. The plotted moments were 
obtained with the full data sample but the center-of-mass rapidity 
restricted to $|y|<$ 0.5, 1, 1.5 and 2 from top to bottom of Fig.~1,
respectively.

The increase of $\rD F_2$ with decreasing $-\ln(\DpT/($GeV/$c))$
for all rapidity cuts and both charge combinations 
in the region of $\DpT$ 
larger than $\sim 1$GeV/$c$ (negative $-\ln\DpT$) can be attributed 
to the conservation of transverse momentum: a particle of so large a 
$\vec p_\rT$ 
is usually accompanied by a number of particles of the opposite transverse 
momentum, so that these particles differ from the former one by large
$\DpT$. Such pairs contribute to the large correlation observed.

The fast rise of $\rD F_2$ with decreasing $\DpT$ 
(increasing $-\ln(\DpT /($GeV/$c))$ in the region of 
$\DpT\le 1$GeV/c ($-\ln(\DpT)>0$) 
corresponds to what is usually assumed to be due to Bose-Einstein
correlations in pion interferometry (for a review see e.g.\ Ref.~11).
If the observed correlation is indeed purely due to the Bose-Einstein effect 
and if the source of the particles is fully incoherent, then
the disconnected correlation function can be expressed by
the Fourier transform of the spatial density of the source:
\beq
C_2(\Delta p) = 1 + |\tilde\rho(\Delta p)|^2.
\eeq
If the transverse part of the source distribution $\rho_\rT(\vec{r}_\rT)$
follows a power law $r_\rT^{-2+\gamma_\rM}$ in some region of $r_\rT$, then
the square of its Fourier transform is governed by the power law 
$\DpT^{-2\gamma_\rM}$ in the corresponding region of the conjugate variable 
$\DpT$. Hence$^5$, we can expect a similar power-law behavior of 
$\rD F_2$:
\beq
\rD F_2 = 1 + a(\DpT)^{-\f}.
\label{powerfit}
\eeq

\begin{table}
\begin{center}
\parbox{4.2in}{
\footnotesize Table 1: Comparison of parameters of the fit of the 
dependence of $\rD F_2$ on $\DpT$ by Eq.~\ref{powerfit} for various 
rapidity windows, in the range indicated by the full line in Fig.~1.
}
\vs 0.3cm
\begin{tabular}{|r|c|c|c|}
\hline
\multicolumn{4}{|c|}{charge combination: $--$} \\ \hline 
rapidity cut & $\f$ & $a\times 10^2$ & $ \chi^2/NDF$ \\ \hline
$|y|<0.5$ & 0.83$\pm$0.12 &   4.6$\pm$1.2  &  11.6/13 \\ \hline
   $<1.0$ & 0.87$\pm$0.10 &   2.8$\pm$0.6  &  21.7/13 \\ \hline
   $<1.5$ & 0.92$\pm$0.10 &   1.9$\pm$0.4  &  27.7/13 \\ \hline
   $<2.0$ & 0.92$\pm$0.10 &   1.5$\pm$0.4  &  34.0/13 \\ \hline
\multicolumn{4}{|c|}{charge combination: $++$} \\ \hline 
rapidity cut & $\f$ & $a\times 10^2$ & $ \chi^2/NDF$ \\ \hline
$|y|<0.5$ & 0.75$\pm$0.10 &   4.7$\pm$0.9  &  22.9/13 \\ \hline
   $<1.0$ & 0.84$\pm$0.09 &   2.4$\pm$0.5  &  15.7/13 \\ \hline
   $<1.5$ & 0.95$\pm$0.12 &   1.3$\pm$0.4  &  18.1/13 \\ \hline
   $<2.0$ & 0.93$\pm$0.16 &   1.0$\pm$0.4  &  23.6/13 \\ \hline
\end{tabular}
\end{center}
\end{table}
 
In Fig.~1 we show the fits of the data with Eq.~\ref{powerfit} as solid lines.
The $\DpT$ region in which momentum conservation has a substantial influence on
the correlation is excluded from the fit. So, only data with $\D p<1$ GeV/$c$
$(-\ln\DpT>0)$ are used.
The parameters obtained in the fits are given in Table~1. In all cases,
the  value of $\f$ agrees (within 2 standard deviations) with the value 
$2\gamma_\rM \approx 0.74$ obtained by Bia\l{}as and Peschanski$^1$. 
The agreement (and the quality of the fit) is better for the smaller than
for the bigger $y$-intervals. This can be understood from the fact that in 
the case of a broad rapidity range, the Bose-Einstein correlations are partly 
washed out due to the large difference possible in the longitudinal momentum.

The parameter $a$ governing the overall value of the correlation 
rises with decreasing size of the rapidity window. This observation
is consistent with the
Bose-Einstein interpretation of the correlation: for a finite longitudinal
size of the source the Bose-Einstein correlation increases when the distance
between the particles in longitudinal momentum decreases.

We conclude that a structure surprisingly 
similar to the one predicted in Ref.~1 for dipole emission can be seen
in our hadron-hadron data.

\section{Conclusions}
We have measured the second-order differential factorial moments $\rD F_2$ 
in the difference of the transverse momentum $\DpT$ in multiparticle production 
at 250GeV/$c$. $\rD F_2$ rises with decreasing $\DpT$ for like-sign particle 
combinations. For narrow rapidity windows the $\DpT$ dependence of $\rD F_2$ 
is fitted quite closely by a simple power law. If the rise of $\rD F_2$ is 
assumed to occur due to the Bose-Einstein correlation, this relationship 
indicates a power-law structure in the transverse-size distribution of the 
source.  For all rapidity windows the slope of that power-law 
dependence is in agreement with the value predicted in Ref.~1 for the 
emission of color dipoles in onium-onium collisions. 
\vspace{0.5cm} 

\noindent
{\Large\bf Acknowledgments}
\vspace{0.2cm}\\
I would like to express my thanks to W.~Kittel for numerous discussions
and careful reading the manuscript. Discussions with A.~Bia\l{}as and
R.~Peschanski are kindly acknowledged. This work was partly supported 
by the Polish KBN grant No.\ 2~P03B~083~08 and by the Polish-German 
Collaboration Foundation grant FWPN no.\ 1441/LN/94.

The EHS/NA22 Collaboration acknowledges early contributions to its experiment
from the III. Physikalisches Institut B, RWTH Aachen, Germany, the 
DESY-Institut f\"ur Hochenergiephysik, Berlin-Zeuthen, Germany, the 
Institute for High Energy Physics, Protvino, Russia,
the Department of High Energy Physics, Helsinki University, Finland,
and the University of Warsaw and Institute of Nuclear Problems, Poland.
This work is part of the research program of the ``Stichting
voor Fundamenteel Onderzoek der Materie (FOM)", which is financially
supported by the ``Nederlandse Organisatie voor Wetenschappelijk Onderzoek
(NWO)". We further thank NWO for support of this project within the program
for subsistence to the former Soviet Union (07-13-038). 

\newpage
\noindent
{\Large \bf References}
\vspace{0.3cm}\\
\begin{tabular}{rl}
1. & A. Bia\l as and R. Peschanski, Phys.\ Lett.\ B355 (1995) 301
\\
2. & \parbox[t]{10.9cm}{
A.H. Mueller, Nucl. Phys. B415 (1994) 373; B437 (1994) 471;
A.H. Mueller, B. Patel, Nucl. Phys. B425 (1994) 471
}
\\
3. & \parbox[t]{10.9cm}{  
A. Bia\l as and R. Peschanski, Nucl. Phys. B273 (1986) 703;
ibid. B308 (1988) 857; for a recent review see: 
E.A. De Wolf, I. Dremin and W. Kittel,  Phys. Reports 270 (1996) 1
}
\\
4. & R. Hanbury-Brown, R.Q. Twiss, Nature 178 (1956) 1046
\\
5. & A. Bia\l as, Acta Phys. Pol. B23 (1992) 561
\\
6. & H.C. Eggers et al.,  Phys. Rev. D48 (1993) 2040
\\
7. &  \parbox[t]{10.9cm}{ 
P. Lipa and H.C. Eggers, Phys.\ Rev.\ D51 (1995) 2138;
Proceedings of the Cracow Workshop on Multiparticle Interactions
``Soft Physics and Fluctuations'', Cracow, 4th-7th May 1993, ed.\ A.~Bia\l{}as,
K.~Fia\l{}kowski, K.~Zalewski, R.C.~Hwa,
(World Scientific, Singapore 1994), p.~93
}
\\
8. & \parbox[t]{10.9cm}{ 
M. Charlet, Star-integrals in NA22, in: Proc. XXIII Int. Symp. on
Multiparticle Dynamics, Aspen 1993, ed. R.C. Hwa (World Scientific,
Singapore, 1994) p.302; M. Charlet, Ph.D. Thesis, University of Nijmegen (1994)
}
\\
9. & M. Adamus et al. (NA22), {Z. Phys.} C32 (1986) 475
\\
10. & M. Adamus et al. (NA22), {Z. Phys.} C39 (1988) 311
\\
11. &  \parbox[t]{11.3cm}{
D.H. Boal, C.-K. Gelbke, B.K. Jennings: Rev.Mod.Phys.\ 62 (1990) 553
}
\end{tabular}

\end{document}

\begin{document}

document}